\documentstyle[aps,prl,multicol]{revtex}

\begin{document}

\draft
\title
{Exchange-Correlation Hole in Polarized Insulators:  \\ Implications
for the Microscopic Functional Theory of Dielectrics}

\author{Gerardo Ortiz$^{a,c}$, Ivo Souza$^{a,b}$, and Richard M. 
Martin$^{a,b}$}

\address{$^a$Department of Physics and $^b$Materials Research
Laboratory, University of Illinois, Urbana, Illinois 61801 }

\address{$^c$Theoretical Division, Los Alamos National Laboratory,
P.O. Box 1663, Los Alamos, NM 87545 }

\date{\today}
\maketitle
\begin{abstract}
We present a simple and direct proof that the exchange-correlation 
hole, and therefore the exchange-correlation energy, in a polarized 
insulator is not determined by the bulk density alone. It is uniquely 
characterized by the density and the macroscopic electric polarization 
of the dielectric medium. 
\end{abstract}

\pacs{PACS numbers: 71.10.-w, 71.15.Mb, 77.84.-s}

\begin{multicols}{2}

\columnseprule 0pt

\narrowtext

In a famous paper in 1964 Hohenberg and Kohn\cite{hoh64} established 
the basic principle of density functional theory (DFT): any property 
of an interacting time-reversal invariant many-particle system is 
uniquely determined as a functional of its ground state density 
$n_0({\bf r})$. The proof rests upon the demonstration that a given 
$n_0({\bf r})$ is consistent with only {\it one} external scalar 
potential $v_{ext}({\bf r})$ (except for an arbitrary constant), which 
determines the entire Hamiltonian since kinetic and 
particle-interaction ($V_{e-e}$) terms are known from the fundamental 
constants $\hbar$, $m_e$, and $e$. Together with the Kohn-Sham 
(KS)\cite{koh65} ansatz that the interacting ground state density can 
be represented by non-interacting fermions in an effective local 
potential, this has provided the foundation for much of the current 
theoretical calculations in atomic, nuclear and condensed matter 
physics\cite{dre90}. 

In the past DFT has often been applied uncritically to condensed 
matter, assuming implicitly that the properties of bulk matter in the 
thermodynamic limit are functionals only of the density in the bulk. 
However, the Hohenberg-Kohn (HK) proofs are only for the entire 
density (including the surface) and they provide no guarantee that the 
functionals of the density are well-defined in the thermodynamic 
limit. In particular, the use of Born-von Karman (BvK) boundary 
conditions may introduce additional considerations not included in 
their original analysis. Although it is well-known that the long-range 
Coulomb terms must be treated carefully and are not absolutely 
convergent in the thermodynamic limit, it has been implicitly assumed 
that the key quantity describing many-body effects in the KS approach, 
the exchange-correlation (x-c) energy $E_{xc}$, is an intrinsic 
functional of the bulk density. Recently, however, Gonze, Ghosez, and 
Godby (GGG)\cite{gon95} have argued that in an insulator $E_{xc}$ must 
also be a function of the macroscopic (average) electric polarization 
in the limit of an infinite system. This has led to much controversy 
and proposals concerning the nature of exchange and 
correlation\cite{res96,MO96,MO97} and the KS 
functional\cite{gon97,van97} in insulators.  

In this Letter we provide a simple direct demonstration that the x-c 
hole in a polarized insulating crystal is modified in a manner {\it 
not described solely by the density in the bulk region of the 
crystal}, even though the hole around an electron in the bulk is 
localized to the bulk region. We show that the hole is determined only 
if the bulk macroscopic electric polarization is specified in addition 
to the bulk density. This can be interpreted as the ``polarization'' 
of the x-c hole\cite{MO96,MO97}, providing the physical mechanism 
leading to the generalized form of the KS equations required to have 
an in-principle-exact description of polarized extended 
matter\cite{gon95,MO97}.  

In a dielectric with a macroscopic electric field ${\bf E_{\sl mac}}$, 
the energy can be expressed in terms of the polarization\cite{lan60}. 
In the case of a macroscopic region of a periodic crystal of volume 
$\Omega$, external effects can be incorporated\cite{MO96,MO97} by 
terms of the form ($\hbar=m_e=-e=1$ in a.u.)
\begin{equation}
E_{ext} = - \Omega \ {\bf E_{\sl mac}} \cdot {\bf P}_{\sl mac} + 
\int_{\Omega} d{{\bf r}} \ v_{ext}({\bf r}) \ n({\bf r}) \ ,
\label{eq:E-gen} 
\end{equation}
where the field couples to the macroscopic polarization ${\bf P}_{\sl 
mac}$, and the second term is explicitly restricted to only the 
periodic part of the external potential $v_{ext}({\bf r})$ coupled to 
the electronic (number) density $n({\bf r})$, which is periodic. The 
key point for our purposes is that ${\bf P}_{\sl mac}$ is {\it not 
directly} determined by the density in $\Omega$. Although variations 
in the polarization density field are related to $n({\bf r})$ by ${\bf 
\nabla} \cdot {\bf P}({\bf r}) = + n({\bf r})$, its average value 
${\bf P}_{\sl mac}$ can be found only as a change from a reference 
polarization ${\bf P}^0_{\sl mac}$, in terms of an integrated 
polarization current that flows {\it through} the volume 
$\Omega$\cite{mar74}. This result for ${\bf P}_{\sl mac}$ is a 
consequence of the BvK boundary conditions used to describe the 
extended system\cite{ort94b}; it is not needed in finite systems and 
was not considered in the original analysis of HK who, implicitly, 
assumed open (vanishing) boundary conditions. Thus, the bulk density 
$n({\bf r})$ and changes in polarization $\delta {\bf P}_{\sl mac} = 
{\bf P}_{\sl mac} - {\bf P}^0_{\sl mac}$ are independent 
experimentally measurable quantities. Furthermore, each is uniquely 
determined by the {\it bulk} ground state wave function, with $\delta 
{\bf P}_{\sl mac}$ given by its Berry's geometric 
phase\cite{kin93,res94,ort94b}.

The case of zero macroscopic field, ${\bf E_{\sl mac}} = {\bf 0}$, is 
of special interest because it can be considered the standard 
reference state of the insulator: then its total energy is not {\it 
explicitly} dependent upon ${\bf P}_{\sl mac}$\cite{MO97}, and the 
original analysis of HK is sufficient to conclude that all bulk 
properties must be {\it functionals} only of the bulk density $n({\bf 
r})$; in particular, this includes the ground state spontaneous 
polarization ${\bf P}^0_{\sl mac}$. On the other hand, if ${\bf E_{\sl 
mac}}$ is not zero, the same reasoning requires that all physical 
quantities must be functions of $\delta {\bf P}_{\sl mac}$ as well as 
functionals of the bulk density. Therefore, the intrinsic description 
of a dielectric material in the presence of an electric field requires 
generalization of the HK theorems\cite{MO97}.  

In the KS approach, $E_{xc}$ is defined to be the difference between 
the internal energy of the interacting many-body system and that of a 
non-interacting system of electrons of the same density $n({\bf r})$. 
The generalization for a polarized system requires that the 
non-interacting system also has the same zero-field polarization 
${\bf P}^0_{\sl mac}$ and change $\delta {\bf P}_{\sl mac}$ in the 
presence of a field. As shown in Ref. \cite{MO97}, $E_{xc}$ is related 
to the interacting N-particle wave function $\Psi$ by a generalized 
constrained Levy formulation:

\begin{eqnarray}
E_{xc}[n; {\bf P}_{\sl mac}] &=& \hspace{-0.1cm}
\min_{\Psi \rightarrow n({\bf r}), {\bf P}_{\mbox{\footnotesize {\sl 
mac}}}} \ \langle \Psi| -\frac{1}{2} \sum_{i=1}^{\rm N} \nabla_i^2 + 
V_{e-e} | \Psi \rangle \nonumber \\
&& - T_{ind}[n;{\bf P}_{\sl mac}] - E_H [n]  \ ,
\label{eq:levy-Exc-n-p}
\end{eqnarray}
where $T_{ind}$ is the kinetic energy of independent fermions and 
$E_H$ is the Coulomb Hartree energy\cite{MO97}. This may also be 
expressed in terms of the Coulomb interaction between the electron 
density at point ${\bf r}$ and the density at ${\bf r}+{\bf u}$ of the 
x-c hole $n_{xc}({\bf r},{\bf r} +{\bf u})$ defined by a coupling 
constant integration\cite{dre90} over all values of the interaction 
$\lambda$ from $\lambda=0$ to its full strength $\lambda=1$, 

\begin{eqnarray}
E_{xc} &=& \frac{1}{2} \int_{\Omega} d{{\bf r}} \ n({\bf 
r}) \int_{\Omega} d{{\bf u}} \ \frac{1}{|{\bf u}|} n_{xc}({\bf r}, 
{\bf r} +{\bf u}) \nonumber \\
                   &=& \frac{{\rm N}}{2} \int_{\Omega} d{{\bf u}} \ 
\frac{1}{|{\bf u}|} \left< n_{xc} ({\bf u}) \right> \ ,
\end{eqnarray}
where $\left< n_{xc}({\bf u}) \right>$ is the density-weighted average 
of the x-c hole. Equivalently, $E_{xc}$ can be written in terms of the 
coupling constant averaged (symmetric) pair correlation function 
$g^{\lambda}$ through the relation $n_{xc}({\bf r}, {\bf r} +{\bf u}) 
= \int_0^1 d\lambda \left[ g^{\lambda}({\bf r},{\bf r} +{\bf u}) - 1 
\right] n({\bf r} +{\bf u})$.  

The energy $E_{xc}/$N can be interpreted as the electrostatic 
interaction between an electron of charge $-1$ and the charge 
distribution $-\left< n_{xc}({\bf u}) \right>$, the total charge of 
which is $+1$, as required by a sum rule. Thus $E_{xc}$ is determined 
by the {\it shape} of $\left< n_{xc}({\bf u}) \right>$. The shape is 
also restricted by other considerations, e.g., it is straightforward 
to show that the density-weighted average of $n_{xc}({\bf r},{\bf r} 
+{\bf u})$ over the entire volume of any finite system $\Omega$ must 
be symmetric in ${\bf u}$, i.e., $\left< n_{xc}({\bf u}) \right> = 
\left< n_{xc}({\bf - u}) \right>$. This follows from the fact that 
correlations among electrons are symmetric in particle exchanges. In 
addition, it can be shown that averages over ${\bf r}$ in a cell of a 
periodic region of a material also lead to a symmetric $\left< 
n_{xc}({\bf u}) \right>$ as long as the hole is localized in the 
relative coordinate ${\bf u}$, so that ${\bf r} +{\bf u}$ is also 
limited to the periodic region. This applies to a crystal in an 
electric field, where the (metastable) ground state wave function is 
invariant to lattice translations even though the Hamiltonian is not. 
In this case, however, we note that other averages of the x-c hole, 
such as the simple average over position, $\int_{\Omega} d{{\bf r}} \ 
n_{xc}({\bf r},{\bf r} +{\bf u})$, are not symmetric. 

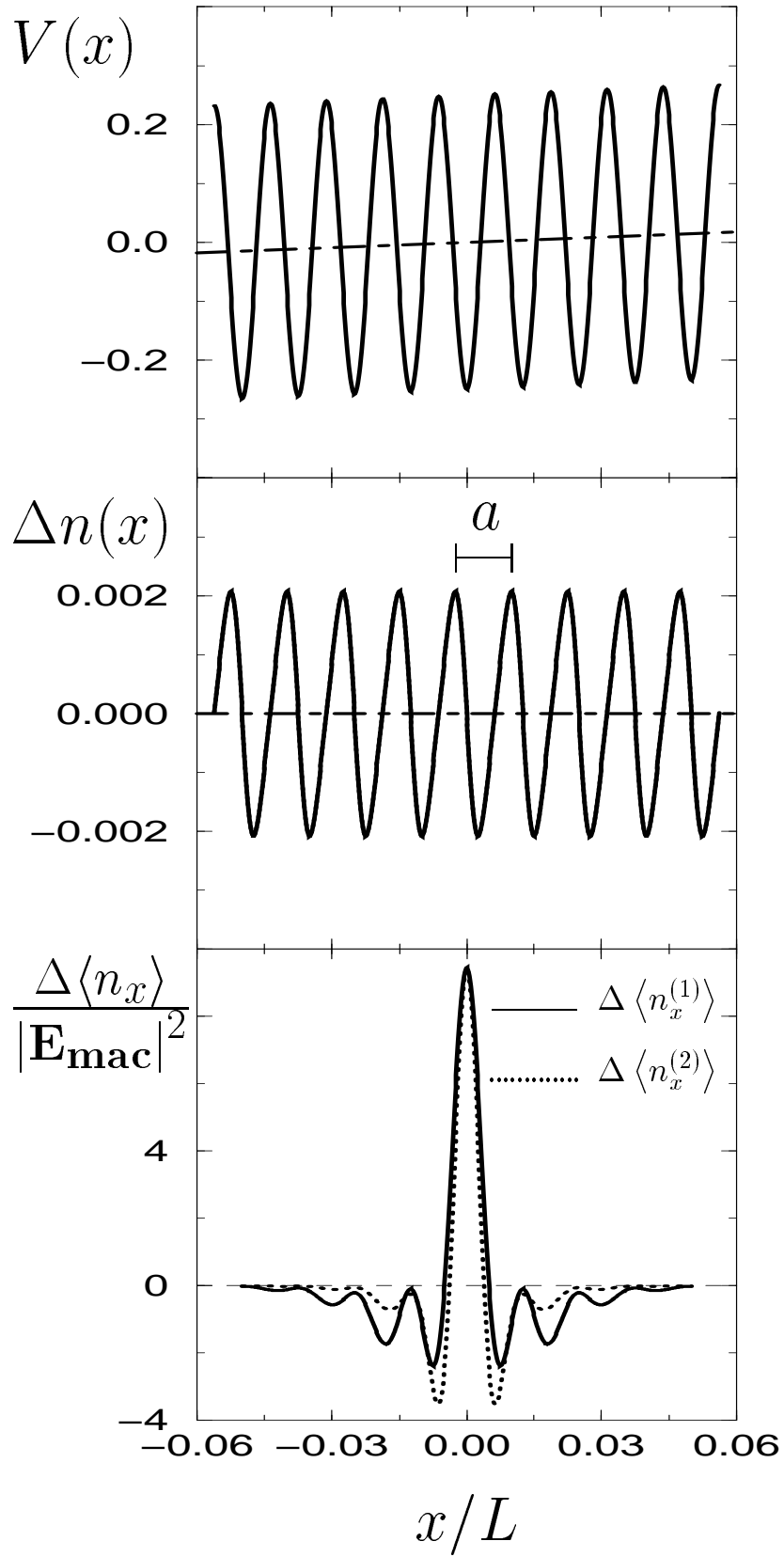
\begin{figure}
\caption{
Results (in a.u.) for a model of a non-polar insulator with the lowest 
band occupied, i.e., 2 electrons/cell. The potential energy (see text) 
is given by $V_0(x)= -V_a \cos(2 \pi x/a)$, $V_a=1/4$, and 
$V_b=V_a/5$, with $a=5 a_0$ ($a_0=\hbar^2/(m_e e^2)$). We show the 
central 8 cells in a supercell of length $L = 80 a$. ($L= 40 a$ gives 
essentially identical results.)
Top panel: $V(x)$; the average slope is ${\bf E}_{\sl mac}=2 \pi 
V_b/L$.
Middle panel: Change in electron density $\Delta n(x) = n(x) - 
n(x)_{V_b = 0}$ in the supercell, and in the fitted system 
(indistinguishable).
Lower panel: Change in the density-weighted average x hole divided by 
$\left|{\bf E}_{\sl mac}\right|^2$ for two different potentials which 
give the same periodic density.}
\label{fig1}
\end{figure}

Now let us consider the case of an insulator in an electric field (or, 
as we will see, the case of a spontaneously polarized insulator). The 
goal is to establish that the x-c hole is both a functional of the 
bulk density and a function of the polarization; for this it is 
sufficient to consider only the non-interacting case where we can 
solve exactly for the exchange (x) hole. It then follows from the fact 
that $E_{xc}$ can be expressed as the coupling constant averaged form, 
for which the non-interacting x hole is one limit, that the same 
conclusions must apply to the exact x-c hole. We consider a 
one-dimensional (1$d$) insulating crystal with potential energy $V(x)= 
V_0(x) + V_{L}(x)$, where $V_0(x)$ is periodic in the crystal lattice 
constant $a$, and $V_{L}(x)$ is an additional long wavelength 
potential with periodicity $L = M a$, $M \gg 1$. In order to mimic a 
constant electric field near $x=0$ we have chosen $V_{L}(x) = V_b 
\left[ \frac{75}{64} \sin(\tilde{x}) - \frac{25}{384} \sin(3 
\tilde{x}) + \frac{3}{640} \sin(5 \tilde{x})\right]$, with $\tilde{x} 
\equiv 2\pi x/L$, which is a symmetric form that is $\propto x + {\cal 
O}(x^7)$. Figure~\ref{fig1} shows the potential energy $V(x)$ and the 
change in density, $\Delta n(x) = n(x) - n(x)_{V_b = 0}$, in the 
region around $x=0$. Although the potential clearly has a linear 
component, the density is very nearly periodic. In the middle panel is 
also shown the change in the periodic density generated by a potential 
$V^{fit}(x)$ which has periodicity $a$ chosen to reproduce the density 
of the central part of the supercell. Since the curves are essentially 
indistinguishable, this illustrates the point made by GGG that the 
same periodic density can be generated by two potentials differing by 
more than a constant.

Consider the density-weighted average x hole around an electron in the 
periodic region around $x=0$ of the supercell. We define this hole to 
be $\langle n^{(1)}_{x} (u) \rangle$ and the change from the zero 
applied field hole to be $\Delta \langle n^{(1)}_{x} (u) \rangle$, 
which is readily expressed in terms of the eigenstates of the 
supercell. We have verified that $\Delta \langle n^{(1)}_{x} (u) 
\rangle$ does not depend upon the unit cell chosen for the average in 
a region around the origin where the electric field is effectively 
constant. Also, the lowest order change in the hole must be quadratic 
in the field strength since the zero-field system has a center of 
inversion symmetry, and we have plotted in Fig.~\ref{fig1} the change 
per unit field {\it squared}. The result is compared 
with the corresponding change for the system with the same density and 
potential $V^{fit}(x)$, which is defined to be $\Delta \langle 
n^{(2)}_{x} (u) \rangle$. It is apparent from the figure that $\Delta 
\langle n^{(1)}_{x} (u) \rangle$ is more extended than $\Delta \langle 
n^{(2)}_{x} (u) \rangle$. Since the two holes are different, we have 
established that the density is not sufficient to determine the x 
hole, and using the reasoning given above we conclude that $E_{xc}$ is 
{\it not determined solely by the bulk periodic density}.  

It is straightforward to prove that the hole is determined by the bulk
density {\it and} the macroscopic polarization. This follows from the 
same reasoning as the original HK arguments except that now the 
external terms in the energy (Eq. (\ref{eq:E-gen})) involve {\it both} 
the bulk $n({\bf r})$ and ${\bf P}_{\sl mac}$. In our present example, 
we have calculated the polarization in the two cases to show 
explicitly that they are different. The average polarization ${\bf 
P}^{(1)}_{\sl mac} = 3.44 \times 10^{-3}$ in cells near the origin of 
the supercell can be found from the density {\it in the supercell} 
using $d {\bf P}(x)/d x = + n(x)$, whereas the polarization ${\bf 
P}^{(2)}_{\sl mac} = 2.11 \times 10^{-3}$ of the system with the 
fitted periodic potential can be found from the Berry's phase 
expressions\cite{kin93,res94}. To our knowledge, this is the first 
explicit demonstration of the polarization dependence of exchange, and 
therefore of the x-c energy.

It is instructive to consider the Clausius-Mossotti limit of 
localized, non-overlapping wave functions in each cell. Then the two 
different ways of calculating ${\bf P}_{\sl mac}$ lead to the same 
answer: the dipole moment per unit length of each isolated unit. The x 
hole is also localized in one cell, so that all properties are 
determined by the density in each cell. In general, however, an 
insulating crystal is not a sum of isolated units, and there is an 
explicit dependence upon ${\bf P}_{\sl mac}$ due to transfer of 
electrons and delocalization of the x hole between cells.

There are other interesting conclusions which follow from the analysis 
of the periodic system in the presence of an electric field, e.g., the 
independent-particle kinetic energy $T_{ind}$ is also a function of 
the polarization\cite{MO97}. The generalized KS functional can be 
written\cite{MO97}
\begin{eqnarray}
\lefteqn{E [n;{\bf P}_{\sl mac}] = E_{ext}[n;{\bf P}_{\sl mac}] }  
\nonumber \\
& & \hspace{1.1cm} + T_{ind}[n;{\bf P}_{\sl mac}] + E_{H} [n] +
E_{xc}[n;{\bf P}_{\sl mac}] \ ,
\label{eq:h-kgen}
\end{eqnarray}
where the first term is given in Eq. (\ref{eq:E-gen}). The variational 
principle requires that $E$ be stationary {\it w.r.t. both} $n({\bf 
r})$ and ${\bf P}_{\sl mac}$, and in terms of the lattice-periodic 
single-particle orbitals $\phi_{i,{\bf k}}$ of the KS approach, the 
Euler-Lagrange equations become\cite{MO97}
\begin{eqnarray}
\lefteqn{\hspace{-1.cm} \left [ -\frac{1}{2}( \nabla + i {\bf k})^2  
+ V_{\sl eff}({\bf r}) + i \ {\bf E}_{\sl eff} 
\cdot \nabla_{\bf k} \right ] \phi_{i,{\bf k}}({\bf r}) } \nonumber \\ 
&& \hspace{3.0cm} =  \ \varepsilon_{i,{\bf k}}({\bf r}) \ 
\phi_{i,{\bf k}}({\bf r})  \ .
\label{eq:KS-var-eq-n-p}
\end{eqnarray}
Here $V_{\sl eff}({\bf r}) = v_{ext}({\bf r}) + \delta (E_{H} + 
E_{xc})/\delta n({\bf r})$ is an ordinary local potential and 
${\bf E}_{\sl eff} = {\bf E}_{\sl mac} - \frac{1}{\Omega} 
\mbox{\boldmath $\partial$} E_{xc}/\mbox{\boldmath $\partial$}{\bf 
P}_{\sl mac}$ is an effective electric field. Accordingly, the 
resulting generalized KS equations are formally the same as the usual 
{\it density-only} ones except that there is an additional term 
involving the derivative of the periodic orbitals with respect to the 
wave vector ${\bf k}$ (where ${\bf k} \in 
\left[-\frac{\pi}{a};\frac{\pi}{a}\right]$ in 1$d$). 

The term $\mbox{\boldmath $\partial$} E_{xc}/\mbox{\boldmath 
$\partial$}{\bf P}_{\sl mac} = - \Omega \ {\bf E}_{\sl xc}$ in the 
generalized KS equations can be identified with an ``x-c electric 
field''\cite{gon95,MO97} in analogy to the actual electric field term. 
It must be emphasized, however, that this is not a true electric 
field; it acts only on electrons and not on other charges; it may be 
longitudinal or transverse and is not related to any density by a 
Poisson-like equation. Since the changes are quadratic in ${\bf 
E}_{\sl mac}$ and thus also in ${\bf P}_{\sl mac}$, it follows that in 
the case of a symmetric crystal ${\bf E}_{\sl xc} \propto {\bf E}_{\sl 
mac}$ for small ${\bf E}_{\sl mac}$. Thus for zero field the system is 
described by the usual {\it density-only} KS equations, but ${\bf 
E}_{\sl xc}$ {\it affects} the dielectric properties. We do not 
attempt to estimate the size of the effect since the present 1$d$ 
model does not fix the full 3$d$ form of the x hole needed to 
calculate the x energy.

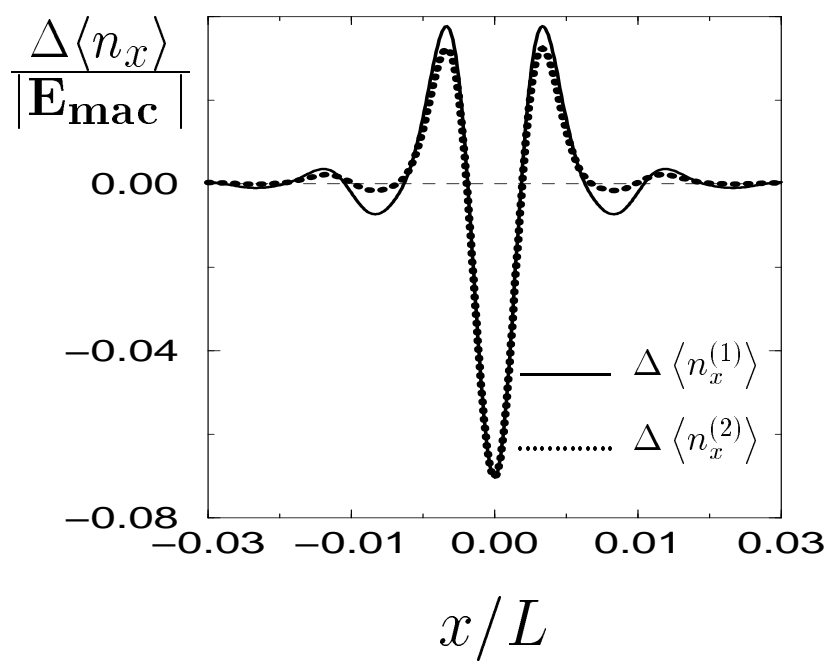
\begin{figure}
\caption{
Same as the lower panel in Fig. 1, but for a polar insulator with 
$V_0(x)= -V_a \left[ \cos(2 \pi x/a) - \frac{1}{4} \sin (4 \pi 
x/a) \right]$, $V_a=1/4$, and $V_b=V_a/20$, plotted per unit field 
${\bf E}_{\sl mac}$. The spontaneous polarization is ${\bf P}^0_{\sl 
mac}=1.321\times 10^{-2}$, while the changes are $\delta {\bf 
P}^{(1)}_{\sl mac}=0.86\times 10^{-3}$, and $\delta {\bf P}^{(2)}_{\sl 
mac}=0.52\times 10^{-3}$. }
\label{fig2}
\end{figure}

If one considers a system with lower symmetry, in general the ground 
state polarization ${\bf P}^0_{\sl mac}$ is non-zero, and there is no 
symmetry reason for ${\bf E}_{\sl xc}$ to vanish. Our calculations 
also provide a microscopic basis for such a field. If we consider a 
1$d$ example in which the crystal potential is non-mirror-symmetric, 
the shape of the x hole changes linearly with the field ${\bf E}_{\sl 
mac}$. An example of the change in the hole per unit field is shown in 
Fig. \ref{fig2}. Since ${\bf P}_{\sl mac}$ varies linearly with ${\bf 
E}_{\sl mac}$, it follows immediately that ${\bf E}_{\sl xc} \neq {\bf 
0}$, even for ${\bf E}_{\sl mac} = {\bf 0}$. Note that the sign of the 
change is relevant, and the positive direction for ${\bf E}_{\sl mac}$ 
in Fig. \ref{fig2} is chosen to be in the same direction as ${\bf 
P}^0_{\sl mac}$. The existence of a spontaneous ${\bf E}_{\sl xc}$ was 
discussed in a recent paper by GGG\cite{gon97}, who described the 
counter-intuitive consequences if one attempts to incorporate it into 
the original density-only KS formulation. Vanderbilt\cite{van97} went 
on to find simple examples where straightforward application 
of the usual KS approach leads to ``ultra-nonlocal dependence upon the 
charge density''. In contrast, we have shown that the generalized KS 
framework provides a simple, direct description in terms of the 
dependence of the x-c hole upon the polarization as well as density.

It is interesting to note that the present theory of polarized 
insulators is formally analogous to the well-known case of spin 
density functionals, in which a spin-polarized system is described by 
KS equations with a spin-dependent effective potential even if there 
is no such magnetic-field-like terms in the original HK system. It is 
necessary to allow for symmetry breaking terms in the KS functional in 
order to describe a broken symmetry state. A general KS formulation 
involving properties other than the density has been developed by 
Jansen\cite{jan91}, who showed that such a theory must in general 
involve operators in the effective Hamiltonian which couple to the 
desired additional variable(s). Jansen's analysis leads to equations 
similar to our Eq. (\ref{eq:levy-Exc-n-p}) in which the effective 
field in principle can be defined in terms of the exact many-body wave 
functions.

In conclusion, we have provided a simple proof that in an insulating 
crystal, the dielectric properties cannot be described solely in terms 
of the periodic bulk density. This leads to a generalized KS approach
needed for all materials with non-zero polarization, in which there is
an effective ``x-c electric field'' ${\bf E}_{\sl xc}$ that can be 
represented as a non-local operator. Our central result is that the 
x-c hole itself is modified in a fundamental way in a polarized 
medium. This was explicitly demonstrated for the x hole in a 
non-interacting system, and it follows from well-known expressions 
involving a coupling constant integration\cite{dre90} that the same 
conclusions apply to the x-c energy $E_{xc}$ in the interacting 
system. Our conclusions have no effect upon work done within 
approximations such as Local Density (LDA) and Generalized Gradient 
(GGA); instead, this work shows that a complete description of 
dielectrics is outside the framework of such approximations.

This work was supported by the NSF grant DMR-94-22496. GO acknowledges 
support from an Oppenheimer fellowship and IS from a PRAXIS XXI 
scholarship. RMM thanks the Aspen Center of Physics for its 
hospitality, and R. Cohen, X. Gonze, W. Kohn, K. Rabe, and D. 
Vanderbilt for stimulating discussions.

\end{multicols}

\end{document}